\documentclass[review, 3p, 12pt]{elsarticle}
\usepackage{amssymb}
\usepackage{amsmath}
\usepackage[colorlinks, allcolors=blue]{hyperref}
\usepackage{tikz}
\usetikzlibrary{
    quantikz2,
    matrix, graphs, positioning,
    fit, quotes, calc,
    arrows.meta, shapes.geometric,
    decorations.markings, decorations.pathreplacing
}
\usepackage{upgreek}
\newcommand*{\upe}{\mathrm{e}}
\newcommand*{\upi}{\mathrm{i}}
\DeclareMathOperator*{\sgn}{sgn}

\graphicspath{{figures/}}

\begin{document}
\begin{frontmatter}

\title{Universal quantum homomorphic encryption based on $(k, n)$-threshold quantum state sharing\tnoteref{t1}}

\tnotetext[t1]{%
    This research was funded by
    the National Natural Science Foundation of China under Grants:
    No.~62401324.
}

\author[1]{Haoyun Zhang\fnref{fn1}}
\author[2]{Yu-Ting Lei\fnref{fn1}}
\author[2]{Xing-bo Pan\corref{cor1}}
\ead{pxingbo@tsinghua.edu.cn}

\cortext[cor1]{Corresponding author}
\fntext[fn1]{These authors contributed equally to this work.}

\affiliation[1]{%
    organization={%
        Department of Physics and Astronomy,
        University of Southern California%
    },%
    city={Los Angeles},%
    state={CA},%
    country={United States}%
}

\affiliation[2]{%
    organization={%
        State Key Laboratory of Low-Dimensional Quantum Physics and Department of Physics,
        Tsinghua University%
    },%
    city={Beijing},%
    country={China}%
}

\begin{abstract}
Quantum homomorphic encryption integrates quantum computing with homomorphic encryption,
which allows calculations to be performed directly on encrypted data without decryption on the server side.
In this paper, we explore distributed quantum homomorphic encryption, focusing on the coordination of
multiple evaluators to achieve evaluation tasks, which not only ensures security but also boosts computational power.
Notably, we propose a $(k,n)$-threshold universal quantum homomorphic
encryption scheme based on quantum state sharing.
Each server is capable of executing a universal gate set,
including the Clifford gates $\{X,Y,Z,H,S,\textsc{cnot}\}$ and a non-Clifford $T$ gate.
The scheme provides that
$k$ evaluation servers chosen from $n$ $(0 < k\leqslant n)$
cooperate to complete the quantum homomorphic encryption
so that the client can get the evaluated plaintext after decryption.
Several concrete examples are presented to provide clarity to our solution.
We also include security analysis, demonstrating its security against eavesdroppers.
\end{abstract}


\end{frontmatter}

\section{Introduction}
\label{sec1}

With the advancement of quantum information, the demand for processing massive amounts of
data has been increasing, making the enhancement of data security and computational capability
an important focus for researchers.
Along with this,
sustained progress has been made in quantum computing
\cite{shorP1994AlgorithmsFor,lanyonB2008ExperimentalQuantum,kimY2023EvidenceFor}
and quantum communication
\cite{bennettC1984QuantumCryptography,pirandolaS2017FundamentalLimits,panD2024TheEvolution},
resulting in many outstanding theoretical
\cite{loH2012Measurement-device-independentQuantum,maX2018Phase-matchingQuantum,liuY2024QuantumIntegrated}
and experimental
\cite{huX2021Long-distanceEntanglement,eckerS2021StrategiesFor,zhangH2022RealizationOf}
achievements.
These advancements have given rise to a new interdisciplinary research area,
quantum homomorphic encryption (QHE)
\cite{broadbentA2015QuantumHomomorphic,dulekY2016QuantumHomomorphic,alagicG2017QuantumFully},
which stands out as particularly noteworthy and attracts much attention.
It enables calculations to be performed on encrypted quantum data without decryption first,
effectively addressing the security challenges associated with massive data processing.

QHE comes from classical homomorphic encryption
\cite{vandijkM2010FullyHomomorphic,brakerskiZ2011FullyHomomorphic,mahadevU2023ClassicalHomomorphic},
particularly a fully homomorphic encryption based on ideal lattices,
which was first proposed by Gentry in 2009 \cite{gentryC2009FullyHomomorphic}.
This development marks a significant breakthrough in the field of cryptography.
In 2012, Rohde \textit{et al.} applied data encryption technology to quantum walking and proposed a scheme for
quantum walking of encrypted data without decryption \cite{rohdeP2012QuantumWalks},
opening up a new path for the combination of quantum
computing and data encryption, which also started the development of QHE.
At present, QHE has made a lot of achievements both theoretically
\cite{liangM2015QuantumFully,ouyangY2020HomomorphicEncryption,chengZ2022ASecure,panX2024CrossQuantum}
and experimentally \cite{fisherK2014QuantumComputing,thamW2020ExperimentalDemonstration,zeunerJ2021ExperimentalQuantum}.

Quantum secret sharing (QSS) is a quantum communication protocol designed to distribute secret information
among multiple participants. Only by acquiring a specific number of shares
can the participants reconstruct the original secret, thereby ensuring the security and privacy of the information.
In 1999, Hillery \textit{et al.} first proposed a quantum secret sharing scheme to encode secret information
into quantum states and distribute them to multiple participants \cite{hilleryM1999QuantumSecret},
which laid an important foundation for the theoretical basis of quantum secret sharing.
Since then, a large number of quantum secret sharing schemes
\cite{fortescueB2012ReducingThe,bellB2014ExperimentalDemonstration,zhouY2018QuantumSecret,liuZ2025ExperimentalDemonstration}
have been proposed.
In 2017, Cao and Ma proposed a $(t, n)$-threshold quantum state sharing scheme using linear equations \cite{caoH2017tN}.
They claim to share a sequence of single-particle
unknown quantum states with $n$ participants and authorize $t$ of them to collaborate to reconstruct the sequence.
In 2019, Chen \textit{et al.} proposed a $(t, n)$-threshold quantum state sharing quantum homomorphic encryption scheme
\cite{chenX2019QuantumHomomorphic} based on Cao and Ma's scheme,
in which $t$ evaluators cooperate to achieve quantum homomorphic encryption with single particle gates as evaluators.
The emergence of numerous $(t, n)$-threshold based quantum protocols
\cite{songX2017tN,ahmadiM2017Relativistic23-threshold,sutradharK2021Enhancedt,liF2022Dynamictn,liF2024ARational}
has advanced the development of distributed
quantum computing, with significant applications in the fields of quantum information security.

Distributed quantum computing assigns quantum computation tasks to
multiple quantum computers or quantum processing units,
allowing them to work collaboratively
to tackle more complex computational problems or solve large-scale quantum algorithms.
The $(k, n)$-threshold system
provides the flexibility when managing evaluators.
In a scenario where the $n$ evaluators provide services to multiple clients simultaneously,
the system allows clients to dynamically select $k$ evaluators from the pool of $n$ based on their
availability, ensuring the computation can proceed without unnecessary delays.
Moreover,
if some evaluators have certain gates that it can carry out more efficiently but lack the
ability to perform others, by allowing the client to choose $k$ evaluators out of the $n$
total, the runtime of the circuit can be optimized according to the specific gates being performed.
This flexibility to select the most suitable subset of evaluators enhances the efficiency
and adaptability of the scheme, accommodating both computational resource constraints
and the diverse capabilities of the evaluators.

In this paper, we propose a universal quantum homomorphic encryption (QHE) scheme
leveraging a $(k, n)$-threshold quantum state sharing mechanism.
This approach enables a client, Alice, to dynamically select $k$ evaluators from $n$ available servers to
collaboratively execute quantum computations on encrypted quantum states.
The process begins with Alice encrypting the qubits using two keys: $\sigma_1$,
which serves as a public key distributed to the $k$ evaluators, and $\sigma_2$,
a private key held exclusively by Alice.
$\sigma_1$ is used during the partial decryption stage, allowing each evaluator to process
the encrypted qubits sequentially and pass the partially decrypted qubits to the next.
The final decryption with $\sigma_2$ completes the computation and ensures the security of the process,
providing enhanced flexibility and security in distributed quantum environments.
While the previous $(k, n)$-threshold scheme supports a limited gate set,
our scheme overcomes the constraint and leverages a universal gate set,
significantly broadening computational possibilities.

\section{Preliminaries}

In this section,
we specify operators of encryption and decryption
used in our scheme.
In addition,
a previous scheme proposed by Chen \textit{et al.} \cite{chenX2019QuantumHomomorphic}
is briefly introduced.

\subsection{Unitary Matrices}

We define two matrices:

\begin{equation}
    U(\gamma) =
    \begin{pmatrix}
    \cos{\gamma} & -\sin{\gamma} \\
    \sin{\gamma} & \cos{\gamma}
    \end{pmatrix},
    \qquad
    E(\beta) =
    \begin{pmatrix}
    1 & 0 \\
    0 & \upe^{\upi\beta}
    \end{pmatrix},
\end{equation}
and these matrices have the property
\begin{equation}
    U(\gamma_1)U(\gamma_2) = U(\gamma_1+\gamma_2), \qquad
    E(\beta_1)E(\beta_2) = E(\beta_1+\beta_2).
\end{equation}
When the $U$ matrix acts on a quantum state $\lvert\psi\rangle = \cos\alpha\lvert 0\rangle + \sin\alpha\lvert 1\rangle$,
it turns to
\begin{equation}
    U(\gamma)\lvert\psi\rangle = \cos(\alpha+\gamma)\lvert0\rangle + \sin(\alpha+\gamma)\lvert 1\rangle.
\end{equation}
Therefore, if we act multiple $U$ matrix on a quantum state,
and the sum of all $\gamma$ in the $U$ matrix is $2\uppi$,
we return to the original state.
This is the underlying basis for encryption and decryption used in our scheme.
Similarly,
The $E$ matrix acting on $\lvert\psi\rangle$ results in
\begin{align}
    E(\beta)\lvert\psi\rangle = \cos\alpha\lvert 0\rangle + \upe^{\upi\beta}\sin\alpha\lvert 1\rangle.
\end{align}

We will later show details on how combinations of the $U$ and $E$ matrix allows us
to achieve more versatile encryption and decryption ability on arbitrary quantum states.

\subsection{Chen's scheme}

Our scheme is mainly inspired by Chen's scheme,
where Alice selects $d$ participants from a group of $n$
($k\le d\le n$),
distributes keys to the $d$ participants,
and passes a set of encrypted quantum states
around them.

It starts with Alice encrypting a set of $L$ qubits,
$\{\lvert\varphi_l\rangle \mid \lvert\varphi_l\rangle=\cos{a_l}\lvert 0\rangle+\sin{a_l}\lvert 1\rangle,l=1,2,\ldots,L\}$
using the $U$ matrix:
\begin{align}
    U(\theta_0)|\varphi_l\rangle
    = \cos(\alpha_l+\theta_0)|0\rangle + \sin(\alpha_l+\theta_0)|1\rangle=
    |\varphi_l^0\rangle,
\end{align}
where $\theta_0 = 2\uppi-\sigma_1-\sigma_2$.
$\sigma_1,\sigma_2\in[0,2\uppi]$ are numbers randomly generated by Alice,
acting as two layers of encryption.
For $i\in\{1,2,\ldots,d\}$,
Alice distributes a decryption key $\theta_i$ to the $i^\textrm{th}$ evaluators.
By Alice's design,
the decryption keys satisfy $\sum_{i=1}^d\theta_i = \sigma_1$, such that $\sigma_1$ will be fully decrypted
after the sequence of qubits have been passed around all $d$ participants.

At the beginning, the encrypted sequence $|\varphi_l^0\rangle$ gets sent to the first participant $A_{i_1}$.
$A_{i_1}$ partially decrypts the qubits
by applying $U(\theta_1)$ on the qubits,
where $\theta_1$ is the private key given by Alice.
Afterwards,
$A_{i_1}$ perform the evaluation operation
$G_1=\prod_{p=1}^{n_1}P_p$ $(n_1\in Z_+$, $P_p \in \{X,Y,Z,H\})$.
Denote the number of $X$, $Y$, $Z$ and $H$ gates in $G_1$
as $x_1$, $y_1$, $z_1$, and $h_1$, respectively,
and let $\eta_1=x_1+z_1+h_1$.
The qubits are now in the state $|\varphi_l^1\rangle=G_1U(\theta_1)|\varphi_l^0\rangle$,
and $A_{i_1}$ send the qubits along with the integer $\eta_1$ to the next evaluator $A_{i_2}$.

$A_{i_2}$ applies $U((-1)^{\eta_1}\theta_2)$ on $|\varphi_l^1\rangle$
to conduct her partial decryption.
The extra $(-1)^{\eta_1}$ term ensures the correct decryption on the qubits
after going through the evaluation operation done by $A_{i_1}$.
$A_{i_2}$ then perform the evaluation operation
$G_2=\prod_{p=1}^{n_2}P_p$ $(n_2\in Z_+$, $P_p \in \{X,Y,Z,H\})$
and calculate $\eta_2=x_2+z_2+h_2$.
The qubits end up in the state $|\varphi_l^2\rangle=G_2U((-1)^{\eta_1}\theta_2)|\varphi_l^1\rangle$,
and $A_{i_2}$ send the qubits and the integer $\eta_1+\eta_2$ to the next evaluator $A_{i_3}$.

For all the following evaluators $A_{i_j}$ $(j<d)$,
$A_{i_j}$ do a partial decryption by applying
$U((-1)^{\eta_1+\eta_2+\cdots+\eta_{j-1}}\theta_j)$
on the received qubits $|\varphi_l^{j-1}\rangle$,
and perform the evaluation operation
$G_j=\prod_{p=1}^{n_j}P_p$ $(n_j\in Z_+$, $P_p \in \{X,Y,Z,H\})$.
The last evaluator $A_{i_d}$ is an exception.
The partial decryption at $A_{i_d}$ follows the standard protocol,
however the allowed evaluation operation at $A_{i_d}$ extends to
$G_d=\prod_{p=1}^{n_d}P_p$, $(n_d\in Z_+$, $P_p \in \{X,Y,Z,H,S,T\})$,
with $s_d$ and $t_d$ denoted as the number of $S$ and $T$ gates performed,
respectively.
$A_{i_d}$ sends $|\varphi_l^d\rangle$,
$\sum_{i=1}^d\eta_i$, $s_d$ and $t_d$ to Alice.

By now, the first layer of encryption $\sigma_1$
has been fully decrypted by the $d$ evaluators.
Alice performs a final decryption operation
\begin{align}
    E(\frac{\uppi s_d}{2}+\frac{\uppi t_d}{4})U((-1)^{\eta_1+\eta_2+\cdots+\eta_{d-1}+\eta_d-1}\sigma_2)E(-\frac{\uppi s_d}{2}-\frac{\uppi t_d}{4})|\varphi_l^d\rangle.
\end{align}
This eliminates $\sigma_2$
and allows Alice to obtain the evaluated plaintext $G_d\cdots G_2G_1|\varphi_l\rangle$.

In Chen's scheme,
the first $d-1$ evaluators are able to perform $X$, $Y$, $Z$, and $H$ gates,
and the last evaluator can perform $T$ and $S$ gates additionally.
However, there are two limitations in Chen's scheme.
First,
none of the $d$ evaluators are able to perform $\textsc{cnot}$ gates on the qubits,
therefore the allowed gates do not form a universal gate set.
Second,
the application of $T$ and $S$ gates can only be done at the final evaluator $A_{i_d}$,
which restricts the freedom of operations Alice can carry out.
In this paper,
we make novel improvements based on their scheme
by using a more robust partial decryption technique.
A universal gate set $\{\textsc{cnot}, X, Y, Z, H, T, S\}$
can be applied at any of the evaluators in our scheme,
greatly expanding the applicability of the $(k, n)$-threshold QHE.

\section{Our scheme}

In this section, we introduce our proposed
universal quantum homomorphic encryption scheme
based on the $(k, n)$-threshold quantum state sharing.
The implementation process of our
scheme is divided into
1) preparation and key generation,
2) encryption phase,
3) partial decryption and collaborative evaluation
and 4) decryption phase.
Figure \ref{fig:concept_diagram} represents a particular system of
our scheme, where Alice is the client, only
responsible for simple key generation and plaintext
encryption and has constrained quantum computing capability.
$\textrm{Bob}_{i_1}$ to $\textrm{Bob}_{i_k}$ are the
evaluation servers responsible
for performing complex quantum calculations,
and they work together to complete a universal
quantum homomorphic encryption.

\begin{figure}[htbp]
    \centering
    \resizebox{.9\textwidth}{!}{\input{figures/diagram.tex}}
    \caption{
        \label{fig:concept_diagram}
        Diagram of $(k, n)$-threshold quantum homomorphic encryption.
        The solid arrows denote the secret keys
        distributed to each evaluation server by Alice,
        and the dashed arrows represent the transmission of quantum ciphertext.
        $k$ evaluation servers $\textrm{Bob}_{i_1}$ to $\textrm{Bob}_{i_k}$
        are selected out of $n$ to collaborate on the evaluation task.
    }
\end{figure}

\subsection{Preparation and key generation}

The scheme starts with Alice preparing her quantum plaintext sequence
$\lvert\varphi\rangle = \{\lvert\varphi_i\rangle \mid \lvert\varphi_i\rangle=\cos{a_i}|0\rangle+\sin{a_i}|1\rangle, i=1,2,\ldots,m\}$,
where $m$ is the length of her plaintext.
She randomly generates $n$ non-zero real elements $b_1, b_2, \ldots,b_n$,
which serve as the public keys.
After that,
She randomly selects $k$ non-zero elements from $b_1, b_2, \ldots,b_n$ to generate the set
$\{b_{i_1},b_{i_2},\ldots,b_{i_k}\}$,
where $1\leq i_1\leq i_2 \leq\cdots\leq i_k\leq n$.
Then, she defines a $k$-variable equation as follows
\begin{align}
b_{i_1}x_{i_1i_2\ldots i_k}+b_{i_2}x_{i_2i_1\ldots i_k}+\cdots+b_{i_k}x_{i_ki_1i_2\ldots i_{k-1}}=1.
\end{align}
She finds a solution $X_{i_1i_2\ldots i_k}(x_f(i_1),x_f(i_2),\ldots,x_f(i_k))$ to the above equation,
where $f(i_j)$ denotes $i_j i_1 i_2 \ldots i_{j-1} i_{j+1} \ldots i_k$,
which is a sequence starting with $i_j$
and followed by the natural order from $i_1$ to $i_{j-1}$ and $i_{j+1}$ to $j_k$.
It should be noted that the solutions to all the equations satisfy that
the same variable corresponds to different values in different equations and every $x_f(i_j)\neq 0$.
Alice randomly selects a non-zero element $\sigma_1$ from the real number field $F$
and defines
\begin{align}
   B_i=\{\sigma_1 b_{i} x_{ij_1j_2\ldots j_{k-1}} \mid i\neq j_1,j_2,\ldots,j_{k-1},1\leq j_1\leq j_2\leq \cdots \leq j_{k-1}\leq n\},
\end{align}
which will be sent to $\textrm{Bob}_i$ as the secret key of him
by secure means,
such as quantum key distribution (QKD)
and quantum direct communication (QSDC).

\subsection{Encryption phase}

Alice generates another random real number $\sigma_2\in F$
as her private key,
and encrypts the plaintext sequence by acting $U(2\uppi-\sigma_1-\sigma_2)$.
This leads the encrypted plaintext to the state
\begin{align}
  |\widetilde{\varphi_i}\rangle=\cos{(2\uppi-\sigma_1-\sigma_2+a_i)}|0\rangle
  +\sin{(2\uppi-\sigma_1-\sigma_2+a_i)}|1\rangle
\end{align}
where $\lvert\widetilde{\varphi_i}\rangle$ is the quantum ciphertext.

\subsection{Partial decryption and collaborative evaluation}

Alice generates and distributes $B_i$ to $n$ evaluation servers
$\textrm{Bob}_1$, $\textrm{Bob}_2$, $\ldots$, $\textrm{Bob}_n$,
and dynamically choose $k$ servers
$\textrm{Bob}_{i_1}$, $\textrm{Bob}_{i_2}$, $\ldots$, $\textrm{Bob}_{i_k}$
from them
to collaborate on the quantum homomorphic encryption.
Each evaluation server executes quantum gates $G_i$ from a set of universal gates,
which includes Clifford gates
$\{X,Y,Z,H,S,\textsc{cnot}\}$ and a
non-Clifford $T$ gate.
The following is the detailed operations
of each evaluation server.

$\pmb{\textrm{Bob}_{i_1}}$:
Alice performs a security check
when she sends the ciphertext sequence to the first evaluation server $\textrm{Bob}_{i_1}$.
That is,
she first randomly selects state vectors from $\{|+\rangle,|-\rangle\}$ and $\{|0\rangle,|1\rangle\}$ bases,
and inserts them into the ciphertext sequence $\{|\widetilde{\varphi_i}\rangle\}$
as check states
to form a new sequence of states $\{|\overline{\varphi_i}\rangle\}$.
After confirming that $\textrm{Bob}_{i_1}$ has received the new sequence,
Alice announces the position of the inserted check states and the corresponding measurement bases.
$\textrm{Bob}_{i_1}$ measures the check states according to
the information published by Alice and sends the results back to Alice.
Then,
Alice calculates the security capacity based on the results,
and in the condition that it is greater than
the preset threshold for channel noise,
the process can continue.
Otherwise,
the sequence is discarded and a new round begins.

After passing the security check,
$\textrm{Bob}_{i_1}$ removes the extra check states
to obtain the original ciphertext sequence $\{|\widetilde{\varphi_i}\rangle\}$.
He performs a partial decryption
using his secret key $B_{i_1}$,
then he can carry out
a set of quantum gates $G_1$
on the ciphertext $\lvert\widetilde{\varphi_i}\rangle$,
obtaining an evaluated state $G_1|\widetilde{\varphi^{1}_i}\rangle$,
where $G_1$ is from the set of universal gates
$\{X$, $Y$, $Z$, $H$, $S$, $T$, $\textsc{cnot}\}$.
He also generates a tuple $Dec_1[\cdot]$,
in which each element serves as the corresponding partial decryption function
for each qubit.
The $\cdot$ will be replaced when the values $B_{i_2}$ are assigned at the next evaluator.
Detailed rules on the generation of these functions are discussed in the next subsection.

Afterwards, $\textrm{Bob}_{i_1}$
randomly inserts check states to generate a new sequence
$\lvert\overline{\varphi^{1}_i}\rangle$
and sends it along with the tuple $Dec_1[\cdot]$
to the second evaluation server $\textrm{Bob}_{i_2}$.

$\pmb{\textrm{Bob}_{i_2}}$:
$\textrm{Bob}_{i_2}$ checks the received sequence
according to the security check method described above.
After the security check is complete
and the original ciphertext sequence $G_1|\widetilde{\varphi^{1}_i}\rangle$ is obtained,
he performs a partial decryption $Dec_1[B_{i_2}]$
using the secret key $B_{i_2}$
and carries out evaluation operators $G_2$,
where $G_2$ is also a combination of gates chosen from the set of universal quantum gates.
He also modifies the tuple $Dec_1[\cdot]$ to $Dec_2[\cdot]$.
As before, He inserts new check states to form a new sequence
and sends it and $Dec_2[\cdot]$ to the next evaluation server $\textrm{Bob}_{i_3}$.
All the following servers will repeat this process,
until the last server $\textrm{Bob}_{i_k}$.

$\pmb{\textrm{Bob}_{i_k}}$:
$\textrm{Bob}_{i_k}$ first performs security checks
to obtain the partially evaluated ciphertext sequence
$G_{k-1}G_{k-2}\cdots G_2G_1\lvert\widetilde{\varphi^1_i}\rangle$
from $\textrm{Bob}_{i_{k-1}}$.
He then performs a partial decryption $Dec_{k-1}[B_{i_k}]$
and carries out his evaluation operators $G_k$.
After he modifies the decryption tuple into $Dec_k[\cdot]$,
he keeps the qubits but send $Dec_k[\cdot]$ to Alice.
Alice will use $Dec_k[\cdot]$ in combination with $\sigma_2$,
which is the private key only she knows,
to calculate a unique decryption tuple $Q^\prime$
and send it back to $\textrm{Bob}_{i_k}$.
Details on how $Q^\prime$ is calculated are shown in following sections.
With $Q^\prime$ and the ciphertext sequence in hand,
$\textrm{Bob}_{i_k}$ performs the last partial decryption
and send the qubits to Alice.
In this way,
Alice's final decryption does not involve any complex quantum operations.

\subsubsection{Evaluation classification}

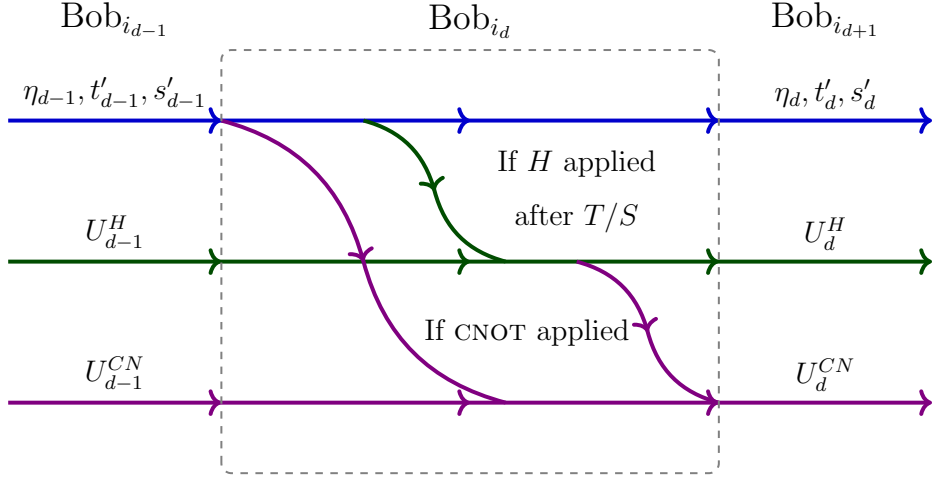
\begin{figure}[htbp]
    \centering
    \resizebox{.75\textwidth}{!}{% !TeX root = ..\main.tex
\begin{tikzpicture}[every path/.style = ultra thick]

\begin{scope}[
    ->,
    decoration = {
        markings,
        mark = at position   3cm with {\arrow{>}},
        mark = at position 6.5cm with {\arrow{>}},
        mark = at position  10cm with {\arrow{>}}
    }
]
    \draw [blue!80!black,postaction={decorate}]   (0,4) -- (13,4);
    \draw [green!30!black,postaction={decorate}]  (0,2) -- (13,2);
    \draw [violet,postaction={decorate}]          (0,0) -- (13,0);
\end{scope}

\begin{scope}[
    decoration = {
        markings,
        mark = at position 0.5 with {\arrow{>}}
    }
]
    \draw [violet,postaction={decorate}]
        (3,4) to [bend left] (5,2) to [bend right] ( 7,0);
    \draw [green!30!black,postaction={decorate}]
        (5,4) to [bend left] (6,3) to [bend right] ( 7,2);
    \draw [violet,postaction={decorate}]
        (8,2) to [bend left] (9,1) to [bend right] (10,0);
\end{scope}

\begin{scope}[every node/.style = above]
    \node at ( 1.5,4) {$\eta_{d-1},t^\prime_{d-1},s^\prime_{d-1}$};
    \node at ( 1.5,2) {$U^H_{d-1}$};
    \node at ( 1.5,0) {$U^{CN}_{d-1}$};
    \node at (11.5,4) {$\eta_d,t^\prime_d,s^\prime_d$};
    \node at (11.5,2) {$U^H_d$};
    \node at (11.5,0) {$U^{CN}_d$};
    \begin{scope}[every node/.append style = {font = \large}]
        \node at ( 1.5,5) {$\textrm{Bob}_{i_{d-1}}$};
        \node at ( 6.5,5) {$\textrm{Bob}_{i_d}$};
        \node at (11.5,5) {$\textrm{Bob}_{i_{d+1}}$};
    \end{scope}
\end{scope}
\node [align=center] at ( 8, 3) {If $H$ applied \\ after $T/S$};
\node                at (7.3,1) {If $\textsc{cnot}$ applied};

\draw [dashed,rounded corners,thick,gray] (3,-1) rectangle (10,5);

\end{tikzpicture}}
    \caption{
        \label{fig:eval_class}
        Evaluation classification for three adjacent servers
        $\textrm{Bob}_{i_{d-1}}$, $\textrm{Bob}_{i_d}$, and $\textrm{Bob}_{i_{d+1}}$.
        The blue line represents only $X$, $Y$, $Z$, $T$ and $S$ gates
        are involved in evaluation.
        In case $H$ gates are applied after $T$ or $S$ gates,
        the partial decryption functions transfers to the green line.
        The violet line is the case
        that $\textsc{cnot}$ gates are applied.
    }
\end{figure}

This part describes the evaluation process in detail for $k$ evaluation servers.
As shown in Figure \ref{fig:eval_class},
the evaluation situation is classified,
where $\textrm{Bob}_{i_{d-1}}$, $\textrm{Bob}_{i_d}$, and $\textrm{Bob}_{i_{d+1}}$
are three adjacent evaluation servers, with $d \leq k-1$.

\textit{First case: X, Y, Z, H gates only.}
The most simple case is $G_d$ purely consists of $X$, $Y$, $Z$, and $H$ gates.
Let $G_d^l$ be the series of single qubit gates performed on the $l^{\textrm{th}}$ qubit,
so the latter becomes
\begin{align}
    \lvert\widetilde{\varphi^d_l}\rangle = G_d^l\lvert\widetilde{\varphi_l^{d}}\rangle^\prime
    = G_d^l U(\sigma_1b_dx_d)\lvert\widetilde{\varphi_l^{d-1}}\rangle,
\end{align}
where $\lvert\widetilde{\varphi_l^{d}}\rangle^\prime$ is the state after partial decryption but before evaluation.
Denote the number of $X$, $Y$, $Z$, and $H$ gates in $G_d^l$
as $x_{d, l}$, $y_{d, l}$, $z_{d, l}$ and $h_{d, l}$.
$\textrm{Bob}_{i_d}$ will send
$\eta_{d, l} = x_{d, l}+z_{d, l}+h_{d, l}+\eta_{d-1, l}$
to $\textrm{Bob}_{i_{d+1}}$.
For example, if $\textrm{Bob}_{i_d}$ applies the series of gate $G_d^l=ZHY$,
he will send $\eta_{d, l}=2+\eta_{d-1, l}$ (2 comes from the one $Z$ and one $H$)
to $\textrm{Bob}_{i_{d+1}}$.
Note that different sets of gates
can be applied on each qubit,
giving them different values of $\eta_{d,l}$,
which are all sent to $\textrm{Bob}_{i_{d+1}}$
in $Dec_d[\cdot]$.

\textit{Second case: T, S gates.}
$\textrm{Bob}_{i_d}$ may apply $T$ or $S$ gates
along with $X$, $Y$, $Z$, and $H$.
Denote the total number of $X$, $Y$, $Z$, $H$, $T$, $S$ gates in
$G_d^l$ as $x_{d, l}, y_{d, l}, z_{d, l}, h_{d, l}, t_{d, l}, s_{d, l}$,
and let $x_{d, l}^T$, $y_{d, l}^T$, $x_{d, l}^S$, $y_{d, l}^S$ be the number of
$X$ and $Y$ gates applied after the first $T$ or $S$ gate, respectively.
At this point, we need to distinguish two subcases
depending on the relative position of $H$ gate with either the $T$ or $S$ gate.

If no $H$ gate was applied after $T$ or $S$,
$\textrm{Bob}_{i_d}$ prepares three integers $\eta_{d, l}$, $t_{d, l}^\prime$, $s_{d, l}^\prime$
as the partial decryption function for $\lvert\widetilde{\varphi_l^{d}}\rangle$.
They are defined as
\begin{gather}
    \eta_{d, l} = x_{d, l}+z_{d, l}+h_{d, l}+\eta_{d-1, l}, \\
    t_{d, l}^\prime =
    \begin{cases}
        (-1)^{x_{d, l}^T + y_{d, l}^T} t_{d, l} & \textrm{if}\ t_{d-1, l}^\prime = 0 \\
        (-1)^{x_{d, l} + y_{d, l}} \sgn(t_{d-1, l}^\prime)
        (|t_{d-1, l}^\prime|+t_{d, l}) & \textrm{if}\ t_{d-1, l}^\prime \neq 0
    \end{cases}, \label{eq:t_neq_0} \\
    s_{d, l}^\prime =
    \begin{cases}
        (-1)^{x_{d, l}^S + y_{d, l}^S} s_{d, l} & \textrm{if}\ s_{d-1, l}^\prime = 0 \\
        (-1)^{x_{d, l} + y_{d, l}} \sgn(s_{d-1, l}^\prime)
        (|s_{d-1, l}^\prime|+s_{d, l}) & \textrm{if}\ s_{d-1, l}^\prime \neq 0
    \end{cases}, \label{eq:s_neq_0}
\end{gather}
where $\sgn(\cdot)$ is the signum function,
and the second cases of Eq.~\eqref{eq:t_neq_0} and \eqref{eq:s_neq_0}
take all $X$ and $Y$ following $\textrm{Bob}_{i_{d-1}}$'s $T$ or $S$ operations
into account.
For example,
assume $\textrm{Bob}_{i_d}$ applied the series of gates $SYXTXTHX$,
and $t_{d-1, l}^\prime$, $s_{d-1, l}^\prime$ received from $\textrm{Bob}_{i_{d-1}}$ are both 0,
the partial decryption function sent to $\textrm{Bob}_{i_{d+1}}$ will be
$\eta_{d, l}=4+\eta_{d-1, l}$,
where the ``4'' comes from three $X$ and one $H$,
$t_{d, l}^\prime=-2$,
because there are two $T$
and the negative sign comes from the three $X$ and $Y$ after the first $T$,
and $s_{d, l}^\prime = 1$,
for the one $S$ and nothing following it.
If $\textrm{Bob}_{i_d}$ received
$t_{d-1, l}^\prime=-1$ and $s_{d-1, l}^\prime=2$,
then the partial decryption function sent to $\textrm{Bob}_{i_{d+1}}$ will be
$\eta_{d, l}=4+\eta_{d-1, l}$,
$t_{d, l}^\prime=(-1)^4\times(-1)\times(|-1|+2)=-3$,
and $s_{d, l}^\prime=(-1)^4\times 1 \times(|2|+1)=3$.

However, if one or more $H$ gates were applied after any $T$ or $S$,
$\textrm{Bob}_{i_d}$ prepares the matrix
\begin{align}
    U^{H}_{G_d^l}(\gamma) = G_d^l\times U(\gamma) \times G_d^{l\dagger}
\end{align}
as the partial decryption function,
where $\gamma$ is a parameter
that will be substituted by real values
at the next evaluator.
For example, if the series of gates $SZHT$ was applied, then
\begin{align}
    U^{H}_{G_d^l}(\gamma) = (S\times Z\times H\times T) \times U(\gamma) \times
    (S\times Z\times H\times T )^{\dagger}.
\end{align}

\textit{Third case: \textsc{cnot} gates.}
$\textrm{Bob}_{i_d}$ may perform \textsc{cnot} gates on a pair of qubits,
$\lvert\widetilde{\varphi_p^{d-1}}\rangle$ and $\lvert\widetilde{\varphi_q^{d-1}}\rangle$.
The $\textsc{cnot}$ gate may be used in combination with any other single qubit gates,
and the series of gates applied on both qubits is denoted as $G_d^{(p, q)}$.
$\textrm{Bob}_{i_{d+1}}$ will need a
$4\times 4$ matrix $U^{CN}_{G_d^{(p, q)}}(\gamma)$
serving as a partial decryption function
to perform the decryption on the pair
$\lvert\widetilde{\varphi_{pq}^d}\rangle$.
To construct $U^{CN}_{G_d^{(p, q)}}(\gamma)$,
$\textrm{Bob}_{i_d}$ first classically computes a
$4\times 4$ matrix $\eta_{G_d^{(p, q)}}$ that represents the composite effect of all gates
applied on $\lvert\widetilde{\varphi_{pq}^{d}}\rangle^\prime$ by himself.
If multiple \textsc{cnot} gates were conducted on non-overlapping pairs of qubits,
a $4\times 4$ matrix shall be computed for each pair.
However, if two or more \textsc{cnot} gates involved overlapping pairs of qubits,
and produced an entangled state of $j$ qubits, $\textrm{Bob}_{i_d}$ must
send the matrix $U^{CN}_{G_d^{(1,2,\ldots,j)}}(\gamma)$ of dimension $2^j \times 2^j$
to $\textrm{Bob}_{i_{d+1}}$,
which is defined as
\begin{equation}
    U^{CN}_{G_d^{(1,2,\ldots,j)}}(\gamma)
    = \eta_{G_d^{(1,2,\ldots,j)}}
    \times
    U(\gamma)^{\otimes j}
    \times
    \eta_{G_d^{(1,2,\ldots,j)}}^\dagger.
\end{equation}

We show an example involving
a 3-qubit state $\lvert\widetilde{\psi^d_{opq}}\rangle$.
If $\textrm{Bob}_{i_d}$ first applies a $\textsc{cnot}$ gate
on the qubits $o$ and $p$
where $p$ is the target qubit,
then passes $o$ through an $X$ gate,
and eventually conducts a $\textsc{cnot}$ on $p$ and $q$
where $q$ is the target,
he calculates $\eta_{G_d^{(o, p, q)}}$ by
\begin{equation}
    \eta_{G_d^{(o, p, q)}} = \textsc{cnot}_{o, p}\times X_o\times \textsc{cnot}_{p, q},
\end{equation}
and the partial decryption function for the 3-qubit state is
\begin{equation}
    U^{CN}_{G_d^{(o, p, q)}}(\gamma) = \eta_{G_d^{(o, p, q)}}\times U(\gamma)\otimes U(\gamma)\otimes U(\gamma) \times \eta_{G_d^{(o, p, q)}}^\dagger.
\end{equation}

After $\textrm{Bob}_{i_d}$ performs his gates
and generates the corresponding partial decryption function for each qubit,
he compiles a tuple $Dec_d[\cdot]$ containing all components in the form of
integers $\eta_{d,l}$, $t_{d,l}^\prime$, $s_{d,l}^\prime$
or matrices $U^{H}_{G_d^l}$, $U^{CN}_{G_d^{(1,2,\ldots,j)}}$.
Next, all $m$ qubits and the tuple $Dec_d[\cdot]$ are sent
to the next evaluation server $\textrm{Bob}_{i_{d+1}}$.

\subsubsection{Partial decryption at following evaluators}

After $\textrm{Bob}_{i_{d+1}}$ receives all $m$ qubits and the tuple $Dec_d[\cdot]$,
if the $l^{\textrm{th}}$
qubit has a corresponding $\eta_{d, l}$, $t_{d, l}^\prime$ and $s_{d, l}^\prime$,
he performs the partial decryption through
\begin{align}\label{eq:pdec_3_integers}
    \lvert\widetilde{\varphi_l^{d+1}}\rangle^\prime
    = E(\frac{\uppi s_{d, l}^\prime}{2} + \frac{\uppi t_{d, l}^\prime}{4})
    U((-1)^{\eta_{d, l}}\sigma_1b_{d+1}x_{d+1})E(-\frac{\uppi s_{d, l}^\prime}{2} - \frac{\uppi t_{d, l}^\prime}{4})
    \lvert\widetilde{\varphi_l^{d}}\rangle.
\end{align}
If the $l^{\textrm{th}}$ qubit has a corresponding matrix $U^{H}_{G_d^l}(\gamma)$, 
then he performs the partial decryption through
\begin{align}
    |\widetilde{\varphi_l^{d+1}}\rangle^\prime =
    U^{H}_{G_d^l}(\sigma_1b_{d+1}x_{d+1})|\widetilde{\varphi_l^{{d}}}\rangle.
\end{align}
If some $j$ qubits are entangled and have a corresponding matrix
$U^{CN}_{G_{d}^{(1, 2, \ldots, j)}}$,
he performs the partial decryption on these $j$ qubits through
\begin{equation}
    \lvert\widetilde{\varphi_{1,2,\ldots,j}^{d+1}}\rangle^\prime
    = U^{CN}_{G_{d}^{(1, 2, \ldots, j)}}(\sigma_1b_{d+1}x_{d+1})
    \lvert\widetilde{\varphi_{1,2,\ldots,j}^{d}}\rangle.
\end{equation}

After $\textrm{Bob}_{i_{d+1}}$ has finished his partial decryption,
he can act any single qubit gates or $\textsc{cnot}$
gates on any of the qubits.
These gates are denoted as $G_{d+1}$.
Afterwards,
he updates the partial decryption function
that he received from the previous evaluator
to $Dec_{d+1}[\cdot]$
and send it to the next evaluation server.

\subsection{Decryption phase}

\begin{figure}[htbp]
    \centering
    \resizebox{.9\textwidth}{!}{% !TeX root = ..\main.tex
\begin{tikzpicture}[every edge/.style={draw,-{Stealth[round]}}]

\matrix [
    nodes in empty cells,
    matrix of math nodes,
    nodes = {anchor = center},
    column sep = 5mm,
    row sep = 3mm,
] (m) {
    \cdots &[2mm] \lvert\widetilde{\varphi^{k-1}}\rangle &[2.6cm]
    \lvert\widetilde{\varphi^k}\rangle & & &
    Q^\prime\lvert\widetilde{\varphi^k}\rangle &[2mm]
    X^aZ^bU(\sigma_2)Q^\prime\lvert\widetilde{\varphi^k}\rangle=G\lvert\varphi\rangle \\
    \cdots & Dec_{k-1}[\cdot] & Dec_k[\cdot] & & & & \\[2mm]
    & & \textrm{find}\ Q & Q^\prime & & & \\
};

\scoped [every edge/.append style=dashed]
\path (m-1-1) edge (m-1-2)
    (m-1-2) edge ["{$G_k Dec_{k-1}[B_{i_k}]$}"{pos=.6,name=gate}] (m-1-3)
    (m-1-3) edge (m-1-6)
    (m-1-6) edge (m-1-7);

\path (m-2-1) edge (m-2-2)
    (m-2-2) edge ["$G_k$"pos=.65] (m-2-3)
    (m-2-3) edge (m-3-3)
    (m-3-3) edge ["$\sigma_2$"] (m-3-4)
    (m-3-4) edge [to path={-| (\tikztotarget)}] (m-1-5);

\node [
    draw, gray,
    dash pattern = on 10pt off 2pt,
    "$\textrm{Bob}_{i_k}$" {font=\large, gray, name=Bik},
    fit = (gate) (m-2-3) (m-1-6)
] (BikBox) {};

\node [
    draw, gray,
    dash pattern = on 10pt off 2pt,
    "$\textrm{Bob}_{i_{k-1}}$" {font=\large, gray},
    fit = (m-1-1) (m-2-2) (m-1-2 |- gate.north)
] {};

\draw [
    gray,
    dash pattern=on 10pt off 2pt
] ([shift={(-3pt,5pt)}] m-3-3.north west)
    -| ([xshift=-3mm] m-1-7.north west |- BikBox.north)
    -- (m-1-7.north east |- BikBox.north)
    |- ([shift={(-3pt,-3pt)}] m-3-3.south west) -- cycle;

\node [font=\large, gray] at (m-1-7 |- Bik) {Alice};

\end{tikzpicture}}
    \caption{
        \label{fig:final_dec}
        Diagram showing Alice's decryption process
        associated with the last evaluation server $\textrm{Bob}_{i_k}$.
        The dashed arrows indicate the transmission direction of qubits,
        and the solid arrows classical information.
    }
\end{figure}
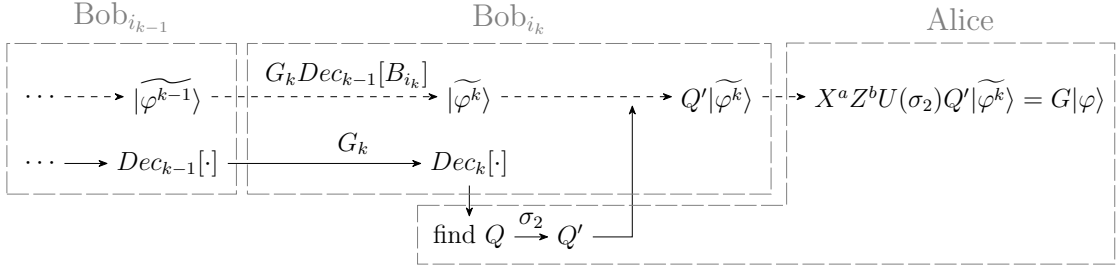

Alice cooperates
with the last evaluator $\textrm{Bob}_{i_k}$
to perform the final decryption in the following way,
as shown in Figure \ref{fig:final_dec}.
She first receives the tuple $Dec_k[\cdot]$ from $\textrm{Bob}_{i_k}$.
If the $l^{\textrm{th}}$ qubit has a corresponding $\eta_{k,l}$, $t_{k,l}^\prime$, and $s_{k,l}^\prime$, the usual decryption
method would be
\begin{align}
    \lvert\varphi_l\rangle = E(\frac{\uppi s_{k, l}^\prime}{2} + \frac{\uppi t_{k, l}^\prime}{4})
    U((-1)^{\eta_{k, l}}\sigma_2)
    E(-\frac{\uppi s_{k, l}^\prime}{2} - \frac{\uppi t_{k, l}^\prime}{4})\lvert\widetilde{\varphi_l^{k}}\rangle.
\end{align}
But since Alice only has the capability of carrying out
the simple $U(\gamma)$, $X$, and $Z$ gates
on the ciphertext (by assumption in any QHE scheme),
she cannot recover the plaintext by getting rid of $\sigma_2$ in this way.
To get around this,
she generates two random one-digit-binary integers for each qubit, $a_l$ and $b_l$,
then classically computes a matrix $Q_l$ as a function of $\sigma_2$ in the following equation
\begin{align}
    X^{a_l} \times Z^{b_l} \times U(\sigma_2) \times Q_l
    = E(\frac{\uppi s_{k, l}^\prime}{2} + \frac{\uppi t_{k, l}^\prime}{4})
    U((-1)^{\eta_{k, l}}\sigma_2)
    E(-\frac{\uppi s_{k, l}^\prime}{2} - \frac{\uppi t_{k, l}^\prime}{4}).
\end{align}
She then plugs the real value of $\sigma_2$ into $Q_l$,
getting a matrix $Q_l^\prime$ consisted of no variables.

If the $l^{\textrm{th}}$ qubit has a partial decryption function $U^H_{G_k^l}(\gamma)$,
Alice calculates $Q_l$ as
\begin{align}
    X^{a_l} \times Z^{b_l} \times U(\sigma_2) \times Q_l = U^H_{G_k^l}(\sigma_2),
\end{align}
and also plugs in the value of $\sigma_2$ to get $Q_l^\prime$.

Similarly, for some $j$ entangled qubits,
the usual decryption method is
\begin{align}
    \lvert\varphi_1\rangle\lvert\varphi_2\rangle\cdots\lvert\varphi_j\rangle= U^{CN}_{G_k^{(1, 2, \ldots, j)}}
    (\sigma_2)\lvert\widetilde{\varphi_1^k}\rangle\lvert\widetilde{\varphi_2^k}\rangle\cdots\lvert\widetilde{\varphi_j^k}\rangle,
\end{align}
but Alice has to find a matrix $Q_{(1,2,\ldots,j)}$ satisfying
\begin{align}
    \bigotimes_{i={1,2,\ldots,j}} \left[ X^{a_i} \times Z^{b_i} \times U(\sigma_2) \right]
    \times
    Q_{(1,2,\ldots,j)}
    = U^{CN}_{G_k^{(1,2,\ldots,j)}}(\sigma_2),
\end{align}
and calculate $Q_{(1,2,\ldots,j)}^\prime$.

She ends up with a tuple $Q^\prime$
and sends it to $\textrm{Bob}_{i_k}$.
Afterwards,
$\textrm{Bob}_{i_k}$ finds a quantum circuit
that can perform each of the partial decryption functions in $Q'$
on the corresponding qubit,
and send the results back to Alice.
The goal of $Q^\prime$ is that after it has been acted,
Alice only needs to perform $X^a Z^b U(\sigma_2)$
to get the evaluated plaintext result $G\lvert\varphi\rangle$,
where $G$ is the accumulation of all conducted evaluation
$G_kG_{k-1}\cdots G_1$.
Since all $a_l$ and $b_l$ are kept in Alice's hand,
the identity of $Q$ is completely hidden from $\textrm{Bob}_{i_k}$.
Thus, even with $Q^\prime$ and $Dec_k[\cdot]$,
he cannot recover $\sigma_2$ and complete the final decryption by himself.
A proof of security is shown in the security analysis section.

\begin{figure}[htbp]
    \resizebox{\textwidth}{!}{% !TeX root = ..\main.tex
\begin{tikzpicture}[
    gate/.style = {draw, dashed, rectangle, minimum height=8mm},
    deckey/.style = {draw, rectangle, minimum height=7mm},
    condition/.style = {draw, diamond, aspect=2.5},
    every matrix/.style = {
        matrix of math nodes,
        matrix anchor = north,
        nodes = {anchor = center},
        row sep = 8mm,
        column sep = -5mm,
    },
    every new ->/.style = {
        -{Stealth[round]}, shorten > = 2pt
    },
    node distance = 1.5cm and 8.5cm,
    vh path/.style = {to path = {|- (\tikztotarget)}},
    hv path/.style = {to path = {-| (\tikztotarget)}},
    vhv path/.style 2 args = {
        to path = {
            -- (\tikztostart |- {$(#1)!.5!(#2)$})
            -| (\tikztotarget)
        }
    },
    vhv path/.default = {\tikztostart}{\tikztotarget},
    hvh path/.style 2 args = {
        to path = {
            -- (\tikztostart -| {$(#1)!.5!(#2)$})
            |- (\tikztotarget)
        }
    },
    hvh path/.default = {\tikztostart}{\tikztotarget}
]

\matrix (B1) {
    & \lvert\widetilde{\varphi}\rangle & \\
    & |[gate]| U(B_{i_1}) & \\
    & |[condition,dashed]| \textrm{Apply gates}\ G_1 & \\
    |[deckey]|\eta_{1,l},t_{1,l}^\prime,s_{1,l}^\prime
    & |[deckey]| U^{CN}_{G_1}
    & |[deckey]| U^H_{G_1} \\
    & |[deckey]| Dec_1[\cdot] & \\
};

\matrix [right=of B1.north] (B2) {
    & \lvert\widetilde{\varphi^1}\rangle & \\
    & |[condition]| Dec_{1}[B_{i_2}] & \\
    |[gate]| E(\gamma_2)U(\theta_2)E(-\gamma_2)
    & |[gate]| U^{CN}_{G_1}(B_{i_2})
    & |[gate]| U^H_{G_1}(B_{i_2}) \\
    & |[condition,dashed]| \textrm{Apply gates}\ G_2 & \\
    |[deckey]|\eta_{2,l},t_{2,l}^\prime,s_{2,l}^\prime
    & |[deckey]| U^{CN}_{G_2}
    & |[deckey]| U^H_{G_2} \\
    & |[deckey]| Dec_2[\cdot] & \\
};

\matrix [right = 6cm of B2.north,
    every matrix/.append style = {row sep=1mm}
] (Bdot) {
    \lvert\widetilde{\varphi^2}\rangle \\[2cm]
    \cdot \\
    \cdot \\
    \cdot \\
    \cdot \\
    \cdot \\
    \cdot \\[2cm]
    \ \\
};

\matrix [right = 7.5cm of Bdot.north] (Bk) {
    & \lvert\widetilde{\varphi^{k-1}}\rangle & &[1.5cm] \\
    & |[condition]| Dec_{k-1}[B_{i_k}] & & \\
    |[gate]| E(\gamma_k)U(\theta_k)E(-\gamma_k)
    & |[gate]| U^{CN}_{G_{k-1}}(B_{i_k})
    & |[gate]| U^H_{G_{k-1}}(B_{i_k}) & \\
    & |[condition,dashed]| \textrm{Apply gates}\ G_k &
    & |[gate]| Q^\prime\lvert\widetilde{\varphi^k}\rangle \\
    |[deckey]|\eta_{k,l},t_{k,l}^\prime,s_{k,l}^\prime
    & |[deckey]| U^{CN}_{G_k}
    & |[deckey]| U^H_{G_k} & \\
    & |[deckey]| Dec_k[\cdot] & & \\
};

\matrix [
    above = 2.3cm of $(B1.north west)!.5!(Bk.north east)$,
    every matrix/.append style = {column sep=1cm}
] (A) {
    \lvert\varphi\rangle
    & |[gate]| U(2\pi-\sigma_1-\sigma_2)
    & \lvert\widetilde{\varphi}\rangle
    & \Big|
    & G\lvert\varphi\rangle
    & |[gate]| X^aZ^bU(\sigma_2)
    & Q^\prime\lvert\widetilde{\varphi^k}\rangle \\
};

\node [deckey,above=12mm of Bk-1-4] (Qp) {$Q^\prime$};
\node [deckey,left=5mm of Qp] (Q) {find $Q$};
\coordinate [below=5mm of Q.south] (p);

\foreach \p / \a / \b in {
    0/B1-3-2/B1-4-2,
    1/B1-4-2/B1-5-2,
    2/B2-2-2/B2-3-2,
    3/B2-3-2/B2-4-2,
    4/B2-4-2/B2-5-2,
    5/B2-5-2/B2-6-2,
    6/Bk-2-2/Bk-3-2,
    7/Bk-3-2/Bk-4-2,
    8/Bk-4-2/Bk-5-2,
    9/Bk-5-2/Bk-6-2%
} \coordinate (p\p) at ($(\a.south)!.5!(\b.north)$);

\graph {
    (A-1-1) -> (A-1-2) -> (A-1-3)
    -> [dashed,vhv path] (B1-1-2) -> (B1-2-2) -> (B1-3-2)
    -- (p0) -> [hv path] {(B1-4-1), (B1-4-2), (B1-4-3)}
    -- [vh path] (p1) -> (B1-5-2);

    (B1-3-2) -> [dashed,hvh path={B1.east}{[xshift=-4mm]B2.west}] (B2-1-2);
    (B1-5-2) -> [hvh path={B1.east}{B2.west}] (B2-2-2);

    (B2-1-2) -> (B2-2-2)
    -- (p2) -> [hv path] {(B2-3-1), (B2-3-2), (B2-3-3)}
    -- [vh path] (p3) -> (B2-4-2)
    -- (p4) -> [hv path] {(B2-5-1), (B2-5-2), (B2-5-3)}
    -- [vh path] (p5) -> (B2-6-2);

    (B2-4-2) -> [dashed,hvh path={B2.east}{[xshift=-4mm]Bdot.west}] (Bdot-1-1);
    (B2-6-2) -> [hvh path={B2.east}{Bdot.west}] (Bdot-2-1);

    (Bdot-7-1) -> [dashed,hvh path={Bdot.east}{[xshift=-4mm]Bk.west}] (Bk-1-2);
    (Bdot-8-1) -> [hvh path={Bdot.east}{Bk.west}] (Bk-2-2);

    (Bk-1-2) -> (Bk-2-2)
    -- (p6) -> [hv path] {(Bk-3-1), (Bk-3-2), (Bk-3-3)}
    -- [vh path] (p7) -> (Bk-4-2)
    -- (p8) -> [hv path] {(Bk-5-1), (Bk-5-2), (Bk-5-3)}
    -- [vh path] (p9) -> (Bk-6-2)
    -> [hv path] (Q) -> ["$\sigma_2$"] (Qp) -> (Bk-4-4)
    -> [dashed,hvh path={Bk-4-4}{[xshift=2cm] Bk-4-4}] (A-1-7)
    -> (A-1-6) -> (A-1-5);

    (Bk-4-2) -> [dashed] (Bk-4-4);
};

\begin{scope}[
    decoration = {brace, amplitude=10},
    every path/.style = decorate,
    every node/.style = {yshift=-8mm, font=\large}
]
    \draw (Bk.south east) --
        node {$\textrm{Bob}_{i_k}$} (Bk.south west);
    \draw (B2.south east |- Bk.south) --
        node {$\textrm{Bob}_{i_2}$} (B2.south west |- Bk.south);
    \draw (B1.south east |- Bk.south) --
        node {$\textrm{Bob}_{i_1}$} (B1.south west |- Bk.south);
    \draw (A.north west) --
        node [yshift=16mm] {Alice} (A.north -| Qp.north east);
\end{scope}

\begin{scope}[gray]
    \draw [dash pattern=on 10pt off 2pt] (B1.north west |- p) -- ([xshift=7mm]Bk.north east |- p);
    \node [anchor=south west,font=\Large] at ([yshift=1cm]B1.north west |- p)  {Client};
    \node [anchor=north west,font=\Large] at ([yshift=-1cm]B1.north west |- p) {Servers};
\end{scope}

\end{tikzpicture}}
    \caption{
        \label{fig:procedure}
        Procedure from Alice's initial encryption to final decryption.
        The dashed boxes indicate operations on quantum states,
        while the solid boxes only involve classical calculations.
        The dashed and solid arrows between entities
        stand for the transmission of quantum and classical information,
        respectively.
        $\gamma_k$ and $\sigma_k$ are described in Eq.~\eqref{eq:pdec_3_integers}.
        Note that the security checks
        in sending the quantum ciphertext sequence
        are omitted in the diagram.
    }
\end{figure}
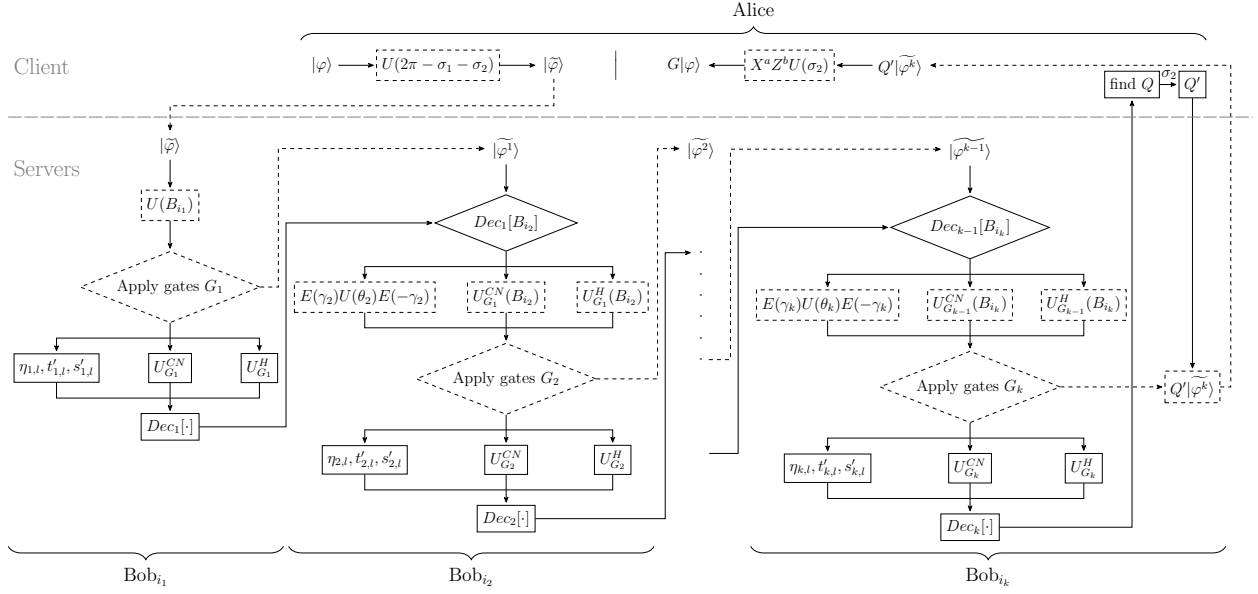

An overall flowchart of the above steps
are summarized in Figure \ref{fig:procedure}.

\section{Examples of our scheme}

In order to explain our proposed scheme more clearly,
three examples are given, including
1) demonstration of multiple $T$, $S$ gates,
2) demonstration of one $\textsc{cnot}$ gate
and 3) demonstration of multiple $\textsc{cnot}$, $T$ and $S$ gates.

\subsection{Example 1: a (3,5)-threshold system applying multiple $T$, $S$ gates}

\begin{figure}[htbp]
    \centering
    \resizebox{.75\textwidth}{!}{\input{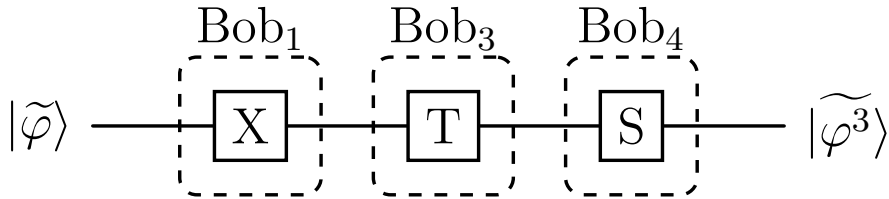}}
    \caption{
        \label{fig:example_1}
        Example of a (3,5)-threshold quantum homomorphic evaluation
        with $T$ and $S$ gates.
    }
\end{figure}

Let's first demonstrate the application of $T$ and $S$ gates on a single qubit
$\lvert\varphi\rangle=\cos\alpha\lvert 0\rangle+\sin\alpha\lvert 1\rangle$.
After Alice applies $U(2\uppi-\sigma_1-\sigma_2)$,
the encrypted state of $\lvert\varphi\rangle$ becomes
\begin{align}
    \lvert\widetilde{\varphi^0}\rangle =
    \begin{pmatrix}
        \cos{(2\uppi-\sigma_1-\sigma_2+\alpha)} \\
        \sin{(2\uppi-\sigma_1-\sigma_2+\alpha)}
    \end{pmatrix}.
\end{align}
Alice distributes partial decryption keys $B_i$
to each evaluator $\textrm{Bob}_i$,
$i\in\{1, 2, 3, 4, 5\}$.
Suppose that she choose
$\textrm{Bob}_1$, $\textrm{Bob}_3$ and $\textrm{Bob}_4$
to conduct an $X$ gate, a $T$ gate and an $S$ gate, respectively,
as shown in Figure \ref{fig:example_1}.

$\textrm{Bob}_1$ performs his partial decryption utilizing $\sigma_1 b_1 x_{134}$
from $B_1$
followed by an $X$ gate,
leaving the encrypted state as
\begin{align}
    \lvert\widetilde{\varphi^1}\rangle = X \times U(\sigma_1 b_1 x_{134}) \times \lvert\widetilde{\varphi^0}\rangle =
    \begin{pmatrix}
        \sin(2\uppi-\sigma_1+\sigma_1 b_1 x_{134}-\sigma_2+\alpha) \\
        \cos(2\uppi-\sigma_1+\sigma_1 b_1 x_{134}-\sigma_2+\alpha)
    \end{pmatrix}.
\end{align}
Since only one $X$ gate is applied,
$\eta_{1} = 1$,
and it is sent to the next evaluator $\textrm{Bob}_3$.
$\textrm{Bob}_3$'s partial decryption is thus
\begin{align}
    \lvert\widetilde{\varphi^2}\rangle^{\prime} = U((-1)^{\eta_1}\sigma_1 b_3 x_{341}) \times \lvert\widetilde{\varphi^1}\rangle =
    \begin{pmatrix}
        \sin(2\uppi-\sigma_1+\sigma_1 b_1 x_{134}+\sigma_1 b_3 x_{341}-\sigma_2+\alpha) \\
        \cos(2\uppi-\sigma_1+\sigma_1 b_1 x_{134}+\sigma_1 b_3 x_{341}-\sigma_2+\alpha)
    \end{pmatrix}.
\end{align}
He then applies a $T$ gate
\begin{align}
    \lvert\widetilde{\varphi^2}\rangle = T \times \lvert\widetilde{\varphi^2}\rangle^{\prime} =
    \begin{pmatrix}
        \sin(2\uppi-\sigma_1+\sigma_1 b_1 x_{134}+\sigma_1 b_3 x_{341}-\sigma_2+\alpha) \\
        \upe^{\upi\frac{\uppi}{4}} \cos(2\uppi-\sigma_1+\sigma_1 b_1 x_{134}+\sigma_1 b_3 x_{341}-\sigma_2+\alpha)
    \end{pmatrix}.
\end{align}
Since the total number of $T$ gate applied is 1 and there are no $X$ or $Y$ gate following,
$t_{2}^{\prime}=t_{2}=1$, and $\eta_{2}$ stays as 1.
$\textrm{Bob}_3$ sends
$t_{2}^{\prime}$ and $\eta_{2}$ to $\textrm{Bob}_4$,
and his partial decryption is
\begin{align}
    \lvert\widetilde{\varphi^3}\rangle^{\prime}
    &= E(\frac{\uppi}{4}) U(-\sigma_1 b_4 x_{413}) E(-\frac{\uppi}{4}) \times
    \lvert\widetilde{\varphi^2}\rangle \\
    &= \begin{pmatrix}
        \sin(2\uppi-\sigma_1+\sigma_1 b_1 x_{134}+\sigma_1 b_3 x_{341}+\sigma_1 b_4 x_{413}-\sigma_2+\alpha) \\
        \upe^{\upi\frac{\uppi}{4}}\cos(2\uppi-\sigma_1+\sigma_1 b_1 x_{134}+\sigma_1 b_3 x_{341}+\sigma_1 b_4 x_{413}-\sigma_2+\alpha) \\
    \end{pmatrix} \\
    &= \begin{pmatrix}
        \sin(2\uppi-\sigma_2+\alpha) \\
        \upe^{\upi\frac{\uppi}{4}}\cos(2\uppi-\sigma_2+\alpha) \\
    \end{pmatrix},
\end{align}
where the last step uses the fact $b_1 x_{134}+b_3 x_{341}+b_4 x_{413}=1$.
After $\textrm{Bob}_4$ applies the $S$ gate, the encrypted state becomes
\begin{align}
    \lvert\widetilde{\varphi^3}\rangle = S \times \lvert\widetilde{\varphi^3}\rangle^{\prime} =
    \begin{pmatrix}
        \sin(2\uppi-\sigma_2+\alpha) \\
        \upe^{\upi\frac{3\uppi}{4}}\cos(2\uppi-\sigma_2+\alpha) \\
    \end{pmatrix}.
\end{align}

Now $\textrm{Bob}_4$ is responsible for assisting Alice with the final decryption.
Since one $S$ gate is applied and there is no $X$ or $Y$ following it,
$s_{3}^{\prime}=s_{3}=1$,
while $t_3^{\prime}=t_2^{\prime}=1$ and $\eta_3=\eta_2=1$,
and these numbers are sent to Alice.
She randomly chooses, for example, $a=0$, $b=1$
and computes the matrix $Q$ in the equation
\begin{equation}
    \begin{aligned}
        X^0 \times Z^1 \times U(\sigma_2) \times Q
        &= E(\frac{3\uppi}{4}) U(-\sigma_2) E(-\frac{3\uppi}{4}) \\
        &= \begin{pmatrix}
            \cos\sigma_2 & -(1+\upi)\frac{\sqrt{2}}{2}\sin\sigma_2 \\
            (1-\upi)\frac{\sqrt{2}}{2}\sin\sigma_2 & \cos{\sigma_2}
        \end{pmatrix}.
    \end{aligned}
\end{equation}
Let's say the random number $\sigma_2$ generated by Alice between 0 and $2\uppi$ is $\uppi/4$.
Alice plugs in the value of $\sigma_2$ into the above equation to get
\begin{align}
    Q^\prime =
    \begin{pmatrix}
        \frac{2-\sqrt{2}}{4} + \upi\frac{\sqrt{2}}{4} & -\frac{2+\sqrt{2}}{4} - \upi\frac{\sqrt{2}}{4} \\
        -\frac{2+\sqrt{2}}{4} + \upi\frac{\sqrt{2}}{4} & \frac{-2+\sqrt{2}}{4} + \upi\frac{\sqrt{2}}{4}
    \end{pmatrix},
\end{align}
and send it to $\textrm{Bob}_4$.
$\textrm{Bob}_4$ carries out the matrix operation
$Q^\prime$ on $\lvert\widetilde{\varphi^3}\rangle$ using a combination of gates,
and returns the encrypted state
\begin{equation}
    Q^\prime \times \lvert\widetilde{\varphi^3}\rangle
    = \begin{pmatrix}
        (\frac{2-\sqrt{2}}{4} + \upi\frac{\sqrt{2}}{4}) \sin(2\uppi-\sigma_2+\alpha) + (-\frac{2+\sqrt{2}}{4} - \upi\frac{\sqrt{2}}{4}) \upe^{\upi\frac{3\uppi}{4}}\cos(2\uppi-\sigma_2+\alpha) \\
        (-\frac{2+\sqrt{2}}{4} + \upi\frac{\sqrt{2}}{4}) \sin(2\uppi-\sigma_2+\alpha) + (\frac{-2+\sqrt{2}}{4} + \upi\frac{\sqrt{2}}{4}) \upe^{\upi\frac{3\uppi}{4}}\cos(2\uppi-\sigma_2+\alpha)
    \end{pmatrix}
\end{equation}
to Alice.
Because $\textrm{Bob}_4$ does not know the values of $a$ and $b$ Alice has chosen,
he cannot extract the value of $\sigma_2$ from $Q^\prime$.
To get the plaintext result of the whole evaluation,
Alice simply performs $X^0\times Z^1\times U(\uppi/4)$ on the returned state
\begin{equation}
    X^0 \times Z^1 \times U(\frac{\uppi}{4}) \times ( Q^\prime \times \lvert\widetilde{\varphi^3}\rangle )
    = \begin{pmatrix}
        \sin\alpha \\
        (-\frac{\sqrt{2}}{2} + \upi\frac{\sqrt{2}}{2}) \cos\alpha
    \end{pmatrix},
\end{equation}
and the result is exactly the evaluation on the plaintext state
$STX\lvert\varphi\rangle$.

\subsection{Example 2: a (2,3)-threshold system applying one $\textsc{cnot}$ gate}

\begin{figure}[htbp]
    \centering
    \resizebox{.75\textwidth}{!}{\input{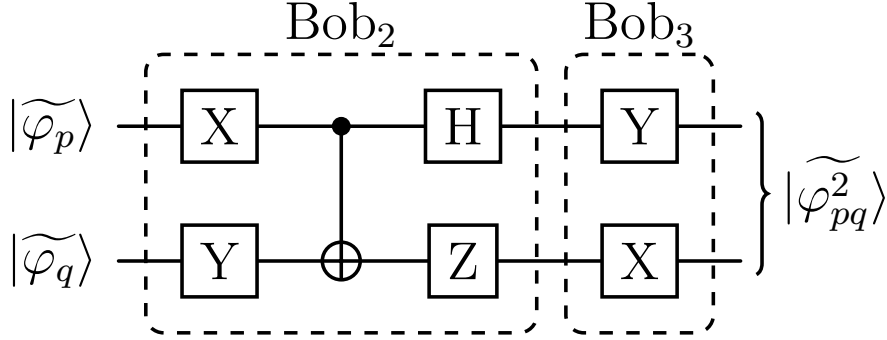}}
    \caption{
        \label{fig:example_2}
        Example of a (2,3)-threshold quantum homomorphic evaluation
        with a $\textsc{cnot}$ gate.
    }
\end{figure}

This example demonstrates the evaluation of one $\textsc{cnot}$ gate
along with other single qubit gates on
a pair of qubits by two evaluators.
Supposed that Alice has plaintext
$\lvert\varphi_p\rangle=\cos{\alpha_p}\lvert 0\rangle+\sin{\alpha_p}\lvert 1\rangle$
and $\lvert\varphi_q\rangle=\cos{\alpha_q}\lvert 0\rangle+\sin{\alpha_q}\lvert 1\rangle$.
She distributes partial decryption keys $B_i$
to the corresponding evaluator $\textrm{Bob}_i$, $i\in\{1,2,3\}$,
and then performs encryption on the two qubits using $U(2\uppi-\sigma_1-\sigma_2)$,
resulting in the encrypted state
\begin{align}
    \lvert\widetilde{\varphi^0_{pq}}\rangle=
    \begin{pmatrix}
        \cos{(2\uppi-\sigma_1-\sigma_2+\alpha_p)}\cos{(2\uppi-\sigma_1-\sigma_2+\alpha_q)} \\
        \cos{(2\uppi-\sigma_1-\sigma_2+\alpha_p)}\sin{(2\uppi-\sigma_1-\sigma_2+\alpha_q)} \\
        \sin{(2\uppi-\sigma_1-\sigma_2+\alpha_p)}\cos{(2\uppi-\sigma_1-\sigma_2+\alpha_q)} \\
        \sin{(2\uppi-\sigma_1-\sigma_2+\alpha_p)}\sin{(2\uppi-\sigma_1-\sigma_2+\alpha_q)} \\
    \end{pmatrix}.
\end{align}

Let's say Alice chooses $\textrm{Bob}_2$ and $\textrm{Bob}_3$ to conduct the evaluation.
$\textrm{Bob}_2$ starts by performing partial decryption and get
\begin{align}
    \lvert\widetilde{\varphi^1_{pq}}\rangle^\prime
    &= U(\sigma_1 b_2 x_{23}) \otimes U(\sigma_1 b_2 x_{23}) \times
    \lvert\widetilde{\varphi_{pq}^{0}}\rangle \\
    &= \begin{pmatrix}
        \cos{(2\uppi-\sigma_1+\sigma_1 b_2 x_{23}-\sigma_2+\alpha_p)}\cos{(2\uppi-\sigma_1+\sigma_1 b_2 x_{23}-\sigma_2+\alpha_q)} \\
        \cos{(2\uppi-\sigma_1+\sigma_1 b_2 x_{23}-\sigma_2+\alpha_p)}\sin{(2\uppi-\sigma_1+\sigma_1 b_2 x_{23}-\sigma_2+\alpha_q)} \\
        \sin{(2\uppi-\sigma_1+\sigma_1 b_2 x_{23}-\sigma_2+\alpha_p)}\cos{(2\uppi-\sigma_1+\sigma_1 b_2 x_{23}-\sigma_2+\alpha_q)} \\
        \sin{(2\uppi-\sigma_1+\sigma_1 b_2 x_{23}-\sigma_2+\alpha_p)}\sin{(2\uppi-\sigma_1+\sigma_1 b_2 x_{23}-\sigma_2+\alpha_q)} \\
    \end{pmatrix}.
\end{align}
He then performs the evaluation
shown in Figure \ref{fig:example_2},
namely an $X$ on $p$
and a $Y$ on $q$,
followed by a $\textsc{cnot}$ where $q$ is the target qubit,
lastly an $H$ on $p$ and a $Z$ on $q$.
After this,
he constructs a partial decryption function
\begin{align}
    U^{CN}_{G_1^{(p, q)}}(\gamma)
    = \eta_{G_1^{(p, q)}}\times U(\gamma)\otimes U(\gamma) \times \eta_{G_1^{(p, q)}}^\dagger,
\end{align}
in which
\begin{align}
    \eta_{G_1^{(p, q)}}
    = H \otimes Z \times \textsc{cnot} \times X \otimes Y.
\end{align}
He sends $U^{CN}_{G_1^{(p, q)}}(\gamma)$
along with the encrypted state $\lvert\widetilde{\varphi^1_{pq}}\rangle$ to $\textrm{Bob}_3$.
$\textrm{Bob}_3$ first performs his partial decryption using $U^{CN}_{G_1^{(p, q)}}(\sigma_1 b_3 x_{32})$,
and the state turns into
\begin{align}
    \lvert\widetilde{\varphi_{pq}^{2}}\rangle^\prime
    = U^{CN}_{G_1^{(p, q)}}(\sigma_1 b_3 x_{32}) \times \lvert\widetilde{\varphi_{pq}^1}\rangle.
\end{align}
Afterwards he performs a $Y$ gate on $p$ and an $X$ gate on $q$,
so $\eta_{G_2^{(p, q)}} = Y \otimes X$. 
The partial decryption function is then transformed into
\begin{align}
    U^{CN}_{G_2^{(p, q)}}(\gamma)
    = \eta_{G_2^{(p, q)}} \times U^{CN}_{G_1^{(p, q)}}(\gamma) \times \eta_{G_2^{(p, q)}}^\dagger.
\end{align}

To finish up,
Alice performs the final decryption together with $\textrm{Bob}_3$,
from whom she receives the decryption function $U^{CN}_{G_2^{(p, q)}}(\gamma)$,
and classically solves for the matrix $Q_{(p, q)}$ in the equation
\begin{align}
    \left[X^{a_1} \times Z^{b_1} \times U(\sigma_2)\right]
    \otimes \left[X^{a_2} \times Z^{b_2} \times U(\sigma_2)\right]
    \times Q_{(p, q)} = U^{CN}_{G_2^{(p, q)}}(\sigma_2).
\end{align}
Assume the values for $\sigma_2$, $a_1$, $b_1$, $a_2$ and $b_2$ are
$3\uppi/2$, 1, 1, 0, and 1, respectively,
then $Q_{(p, q)}^\prime$ is calculated to be
\begin{align}
    Q_{(p, q)}^\prime =
   \begin{pmatrix}
    0 & -1 &  0 & 0 \\
    1 &  0 &  0 & 0 \\
    0 &  0 &  0 & 1 \\
    0 &  0 & -1 & 0
    \end{pmatrix}.
\end{align}
Once $\textrm{Bob}_3$ carries it out using a combination of gates,
the qubits will be returned to Alice,
so she can use simple $U(\sigma_2)$, $X$ and $Z$
to complete the final decryption and get
\begin{equation}
    \begin{pmatrix}
        -\frac{\sqrt{2}}{2} \cos(\alpha_p - \alpha_q) \\
        \frac{\sqrt{2}}{2} \sin(\alpha_p + \alpha_q) \\
        -\frac{\sqrt{2}}{2} \cos(\alpha_p + \alpha_q) \\
        -\frac{\sqrt{2}}{2} \sin(\alpha_p - \alpha_q)
    \end{pmatrix},
\end{equation}
which is exactly the evaluation on the plaintext state,
namely $(Y\otimes X)(H\otimes Z)\textsc{cnot}_{p,q}(X\otimes Y)\lvert\varphi_p\rangle\otimes\lvert\varphi_q\rangle$.

\subsection{Example 3: a (2,3)-threshold system applying multiple $\textsc{cnot}$, $T$ and $S$ gates}

\begin{figure}[htbp]
    \centering
    \resizebox{.75\textwidth}{!}{\input{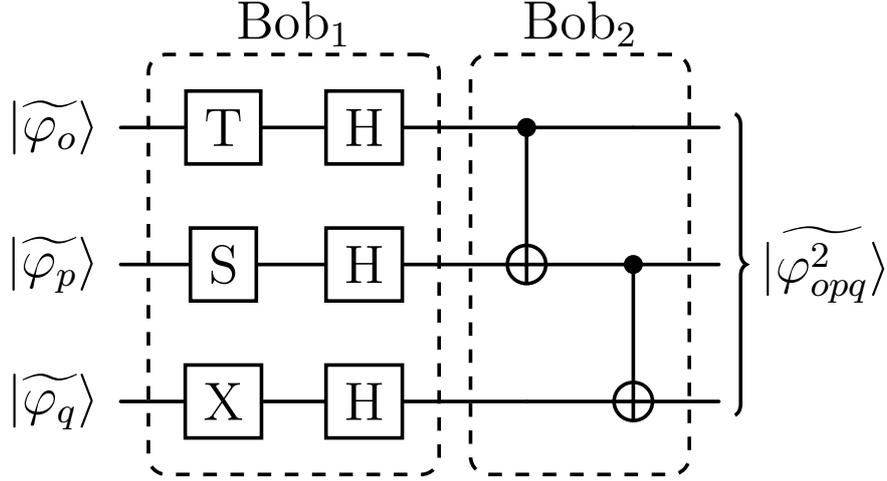}}
    \caption{
        \label{fig:example_3}
        Example of a (2,3)-threshold quantum homomorphic evaluation
        with multiple $\textsc{cnot}$, $T$ and $S$ gates.
    }
\end{figure}

In this example we demonstrate a combination of $\textsc{cnot}$, $T$ and $S$ gate
on three qubits and present a more complicated case.
The sequence of gates applied is shown in Figure \ref{fig:example_3}.
Suppose $\textrm{Bob}_1$ applies $T$ and $H$ gates on the first qubit $o$,
$S$ and $H$ on the second qubit $p$,
and $X$ and $H$ gates on the third qubit $q$.
Then the three qubits are passed on to the next evaluator, $\textrm{Bob}_2$,
who applies one $\textsc{cnot}$ gate on $o$ and $p$,
followed by another $\textsc{cnot}$ on $p$ and $q$.
All calculations up to the partial decryption of $\textrm{Bob}_1$ are
the same as previous example,
so we start from after he has applied all the corresponding gates
on each of the three qubits,
and mainly focus on generating partial decryption functions.
For $o$ and $p$, since both of them had an $H$ gate applied after $T$ or $S$,
their partial decryption functions are generated by
\begin{align}
    U^{H}_{G_1^o}(\gamma) = G_1^o\times U(\gamma) \times G_1^{o\dagger}, \quad
    U^{H}_{G_1^p}(\gamma) = G_1^p\times U(\gamma) \times G_1^{p\dagger},
\end{align}
where
\begin{align}
    G_1^o = H \times T =
    \frac{1}{\sqrt{2}}\begin{pmatrix}
        1 & \upe^{\upi\uppi/4} \\
        1 & -\upe^{\upi\uppi/4}
    \end{pmatrix}, \quad
    G_1^p = H \times S =
    \frac{1}{\sqrt{2}}\begin{pmatrix}
        1 & \upi \\
        1 & -\upi
    \end{pmatrix}.
\end{align}
For the third qubit $q$,
since $X$ and $H$ are applied, $\eta_{1, q}=2$.
$\textrm{Bob}_1$ generates a tuple $Dec_1[\cdot]=\left\{U^H_{G_1^o},U^H_{G_1^p},\eta_{1, q}\right\}$
and sends it to $\textrm{Bob}_2$.
$\textrm{Bob}_2$ plugs in $\sigma_1b_2x_{21}$ into above functions
to conduct the partial decryption,
and then applies two $\textsc{cnot}$ gates.
This entangles the three qubits,
so we find $\eta_{G_2^{(o, p, q)}}$ by
\begin{align}
    \eta_{G_2^{(o, p, q)}} = \textsc{cnot}_{(p, q)} \otimes \textsc{cnot}_{(o, p)},
\end{align}
and the partial decryption function
of the encrypted state $\lvert\widetilde{\varphi_{opq}^{2}}\rangle$
is given by
\begin{align}
    U^{CN}_{G_2^{(o, p, q)}}(\gamma) = \eta_{G_2^{(o, p, q)}}\times U^{H}_{G_1^o}(\gamma) \otimes
    U^{H}_{G_1^p}(\gamma) \otimes U((-1)^{\eta_{1, q}}\gamma) \times \eta_{G_2^{(o, p, q)}}^\dagger.
\end{align}

$\textrm{Bob}_2$ sends $Dec_2[\cdot]=\left\{U^{CN}_{G_2^{(o, p, q)}}\right\}$ to Alice,
and as shown in previous examples,
she calculates the matrix $Q_{(o,p,q)}^\prime$ from
\begin{align}
    \bigotimes_{i={o,p,q}} \left[ X^{a_i} \times Z^{b_i} \times U(\sigma_2) \right]
    \times
    Q_{(o,p,q)}
    = U^{CN}_{G_k^{(o,p,q)}}(\sigma_2),
\end{align}
and completes the final evaluation under the assistance of $\textrm{Bob}_2$.

\section{Security analysis}

In this section, we analyze the security of our scheme.
Alice initially encrypts the qubits using the $U$ matrix,
in which the encryption factors $\sigma_1$ and $\sigma_2$ are random.
When she interacts with the last evaluator to perform the final decryption,
the use of $X^aZ^b$ prevents any eavesdropper from obtaining $\sigma_2$,
thus the security of the plaintext is guaranteed.

To verify the security of encryption with the $U$ matrix,
the density matrix of the single qubit plaintext state
$\lvert\varphi\rangle = \cos\alpha\lvert0\rangle+\sin\alpha\lvert1\rangle$
is
\begin{align}
    \rho=
    \begin{pmatrix}
        \cos^2\alpha & \cos\alpha\sin\alpha \\
        \cos\alpha\sin\alpha & \sin^2\alpha
    \end{pmatrix},
\end{align}
and after encryption with $U(\sigma)$ the density matrix of the resulting state is
\begin{align}
    \rho_{enc} = \frac{1}{4\uppi}\int_{-2\uppi}^{2\uppi}U(\sigma)\rho U(\sigma)^\dagger \mathrm{d}\sigma.
\end{align}
Here, $\sigma$ is entirely equivalent to $2\uppi-\sigma_1-\sigma_2\in[-2\uppi,2\uppi]$
as stated in previous sections of the paper,
since both are random numbers generated by equivalent probability distribution function.
The integral comes from the fact that $\sigma$ can take an infinite number of possible values,
and the result is
\begin{align}
    \rho_{enc} = \frac{1}{2}
    \begin{pmatrix}
        1 & 0 \\
        0 & 1
    \end{pmatrix}
    = \frac{1}{2} I,
\end{align}
where $I$ is the identity matrix,
proving that the state after encryption with the $U$ matrix is a maximally mixed state.

The security of encrypting with $X^aZ^b$ can be proven by similar means.
After $\textrm{Bob}_{i_k}$ finishes his evaluation,
he sends the decryption function $U^\prime(\gamma)$ to Alice,
who calculates $Q$ using $X^a \times Z^b \times U(\gamma) \times Q = U^\prime(\gamma)$,
and plugs the numerical value $\sigma_2$ into $Q$ to get $Q^\prime$
before sending it back to $\textrm{Bob}_{i_k}$.
The purpose of this encryption is therefore
to prevent $\sigma_2$ from leaking to $\textrm{Bob}_{i_k}$,
otherwise he could reconstruct the evaluated plaintext on his own.
The density matrix of $Q = U^{-1}(\gamma)Z^{-b}X^{-a}U^\prime(\gamma)$ is calculated as
\begin{align}
    \rho_Q = \frac{1}{4}\sum_{a,b\in\{0,1\}}[U^{-1}(\gamma)Z^{-b}X^{-a}U^\prime(\gamma)] \times [U^{-1}(\gamma)Z^{-b}X^{-a}U'(\gamma)]^\dagger = \frac{1}{2}I,
\end{align}
which is also a maximally mixed state,
and $\textrm{Bob}_{i_k}$ cannot extract valuable information from it.

The above analysis proves that only Alice knows the value of $\sigma_2$
and has access to the evaluated plaintext.
This eliminates the possibility of internal attacks,
where malicious evaluation servers
may want to steal the plaintext of Alice.
The process of security checks
in sending the sequence of qubits between servers
ensures no external eavesdropper Eve
can gain any encrypted quantum information.
Even if in practical systems
that Eve intercepts some qubits through,
i.e.\ photon number splitting attacks,
she cannot recover the plaintext
without $\sigma_1$ and $\sigma_2$.
Thus, our scheme is secure and feasible.

\section{Conclusion}

Distributed quantum computing plays a crucial role in privacy protection and secure computation.
The universal quantum homomorphic encryption scheme based on $(k, n)$-threshold quantum state sharing
proposed in this paper represents an advancement in the field of distributed quantum computing.
In our scheme, Alice can dynamically select $k$ evaluation servers from a total of $n$ to perform quantum homomorphic encryption based on her specific requirements.
The scheme accounts for all possible evaluation scenarios and categorizes them accordingly.
Initially, Alice encrypts a set of qubits and sends the ciphertext to the first server.
Each subsequent evaluation server executes a set of universal quantum gates as required by Alice.
Once all $k$ servers have completed their evaluations,
Alice performs a final decryption
using her private key,
with assistance from the last server,
to obtain the evaluated plaintext.
The security analysis confirms that Alice's plaintext sequence is never revealed to any server,
and no information is compromised throughout the entire process.
Our scheme is the first to realize $(k, n)$-threshold universal quantum homomorphic encryption in the context of distributed quantum computing.
This approach holds significant potential for future applications in distributed quantum networks,
enabling Alice to leverage a network of evaluation servers for quantum computations
that exceed her individual capabilities.

\bibliographystyle{elsarticle-num}
\bibliography{main}

\end{document}